\documentstyle[twocolumn,prb,aps,epsfig,floats,eqsecnum]{revtex}

\begin{document}
\draft

\twocolumn[\hsize\textwidth\columnwidth\hsize\csname @twocolumnfalse\endcsname

\title{Orbital effect of Cooper pairs on Kondo effect in unconventional superconductors}

\author{Masashige Matsumoto$^1$ and Mikito Koga$^2$}
\address{$^1$Department of Physics, Faculty of Science, Shizuoka University, 836 Oya,
Shizuoka 422-8529, Japan \\
$^2$Department of Physics, Faculty of Education, Shizuoka University, 836 Oya,
Shizuoka 422-8529, Japan}

\date{March 3, 2001}
\maketitle

\begin{abstract}
A new type of Kondo effect peculiar to unconventional superconductors is studied theoretically
by using the Wilson's numerical renormalization group method.
In this case, an angular momentum of a Cooper pair plays an important role in the Kondo effect.
It produces multichannel exchange couplings with a local spin.
Here we focus on a $p_x +i p_y$-wave state which is a full gap system.
The calculated impurity susceptibility shows that the local spin is almost quenched
by the Kondo effect in the strong coupling region ($T_{\rm K}/\Delta \rightarrow \infty$),
while the ground state is always a spin doublet over all the $T_{\rm K}/\Delta$ region.
Here $T_{\rm K}$ and $\Delta$ are the Kondo temperature and the superconducting energy gap,
respectively.
This is different from the $s$-wave pairing case
where the Kondo singlet is realized for large $T_{\rm K}/\Delta$ values.
The strong coupling analysis shows that
the $p_x +i p_y$-wave Cooper pair is connected to the Kondo singlet
via the orbital dynamics of the paired electrons,
generating the spin of the ground state.
This type of Kondo effect reflects the symmetry of the conduction electron system.
\end{abstract}


\vspace{10pt}

]\narrowtext

\newcommand{\br}{{\mbox{\boldmath$r$}}}
\newcommand{\bk}{{\mbox{\boldmath$k$}}}
\newcommand{\sk}{{\mbox{\footnotesize $k$}}}
\newcommand{\bsk}{{\mbox{\footnotesize \boldmath$k$}}}
\newcommand{\bS}{{\mbox{\boldmath$S$}}}
\newcommand{\bd}{{\mbox{\boldmath$d$}}}
\newcommand{\bsigma}{{\mbox{\boldmath$\sigma$}}}

\section{introduction}
The search for variety in the Kondo effect was stimulated by the discovery of the non-Fermi
liquid behavior in heavy fermion materials and by the observation of the two-level system
character in metallic glasses and nanoscale devices.
In these cases, orbital degrees of freedom are very important as well as spin.
\cite{Noziers,Cox98}
Due to incomplete compensation of a local moment,
it displays singular behavior in physical quantities,
differing from the ordinary single-channel Kondo effect.
This Kondo problem has been investigated mainly for metallic systems with no gap
in the density of states.

For a gapped system, the problem of a magnetic impurity in a $s$-wave superconductor has
been investigated for a long time.
\cite{Shiba68,Sakurai,Muller,Matsuura,Jarrell,Satori,Sakai,Yoshioka}
Since the $s$-wave property is described by a single channel, the Kondo effect itself is
understood in the framework of the conventional theory based on the single-channel Kondo model.
On the other hand, no one has paid attention to multichannel properties of $p$-wave and $d$-wave
superconductors derived from angular momenta of their Cooper pairs.
They can exhibit a new type of Kondo effect peculiar to the unconventional superconductivity.

The unconventional superconductivity is very sensitive to impurities and surface boundaries.
It displays peculiar phenomena not observed in the $s$-wave superconductivity.
It is known that scattering objects destroy the unconventional superconductivity
and induce bound states.
\cite{Schmitt-Rink}
The nuclear magnetic resonance (NMR) $T_1^{-1}$ measurement
performed in Zn-doped YBa$_2$Cu$_3$O$_{7-\delta}$
showed a constant $T_1 T$ at low temperatures
\cite{Ishida93}
reflecting finite density of states at lower energies.
As shown by the low temperature scanning tunnel microscope (STM),
a single Zn atom actually generates the low energy bound state.
\cite{Pan}
For the $d_{x^2-y^2}$-wave,
the (1,1,0) surface is regarded as the pair-breaking boundary.
\cite{Hu}
A very recent experiment showed
a zero energy peak in the tunneling conductance
between the (1,1,0) boundary of YBa$_2$Cu$_3$O$_{7-\delta}$ and a normal metal.
\cite{Iguchi}
Thus much attention has been paid to the effects of non-magnetic impurities and static
boundaries in the last decade.

Recently the effect of magnetic impurities in unconventional superconductors has attracted
attention gradually with the development of experimental studies.
According to the NMR experiments,
\cite{Alloul}
a local moment is induced around a Zn site substituted for a Cu site
in YBa$_2$Cu$_3$O$_{7-\delta}$.
It is expected that the observed Kondo behavior is due to this moment.
\cite{Sisson}
Similar behavior is also observed in Li-substituted YBa$_2$Cu$_3$O$_{7-\delta}$.
\cite{Bobroff}
In this case, the local magnetic susceptibility of the Li induced moment shows strong
reduction of the Kondo screening below the superconducting transition temperature.
At present the search for the influence of unconventional superconductivity on the
Kondo effect is still in progress.
We are also interested in the Kondo effect for different types of superconductors
other than such high-$T_{\rm c}$ cuprates.

Several studies of a magnetic impurity in a conventional ($s$-wave) superconductor clarified
how a bound state generated around the impurity depends on the Kondo temperature $T_{\rm K}$
and the superconducting energy gap $\Delta$.
\cite{Matsuura,Jarrell,Satori,Sakai,Yoshioka}
The numerical renormalization group (NRG) study by Satori {\it et al.}
\cite{Satori}
showed that the bound state energy is scaled by $T_{\rm K} / \Delta$
and the interchange of singlet and doublet ground states occurs smoothly.
Even if an energy gap exists in the density of quasiparticle states, a local spin couples with
the quasiparticles antiferromagnetically to form a spin singlet (Kondo singlet) for
$T_{\rm K} / \Delta > 0.3$.
A similar competition was discussed in gapless $d_{x^2 - y^2}$-wave superconductors
such as high $T_{\rm c}$ cuprates.
\cite{Simon}
In this case, the gapless density of states stabilizes the Kondo singlet more favorably
than in the gapped $s$-wave case.
However, it is not certain that a singlet ground state is always realized in all types
of superconductors when the Kondo effect is strong enough.
A Cooper pair in the unconventional superconductivity has an angular momentum and can produce
multichannel scattering of quasiparticles at an impurity.
As well as for metals,
a difference between single channel and multichannel properties in the Kondo
effect can be expected for superconductors.

In order to elucidate the multichannel property, we study the Kondo effect due to
a local spin coupled to a $p_x + i p_y$
(or $d_{x^2 - y^2} + i d_{xy}$)-wave superconducting state.
The $\bd$-vector for the $p_x + i p_y$-wave is given by
$\bd_\bsk = \hat{z} (k_x + i k_y)$ where $\hat{z}$ is a unit vector in the z axis.
Since the density of states has a full gap like the $s$-wave case, we can find the peculiarity
of the unconventional superconductivity in comparison with the $s$-wave.
The property of the $p_x + i p_y$ superconductivity is analogous to that of the A phase
found in the superfluid $^3$He.
A lot of attention has been paid to this superconductivity since the discovery of the
spin triplet superconductor Sr$_2$RuO$_4$.
\cite{Maeno,Ishida98,Rice}
The recent theoretical studies proposed a number of unique physics associated with
the broken time reversal symmetry observed by the $\mu$SR experiment.
\cite{Luke}
In the $p_x + i p_y$-wave superconductor, spontaneous current appears close to the
scattering objects
and it induces a finite magnetic field.
\cite{Matsumoto99,Okuno}
This argument explains the $\mu$SR data well.
\cite{Matsumoto00-1}
If the system induces the spontaneous magnetic field by itself,
a spontaneous Hall effect can be expected.
\cite{Goryo99}
A vortex charging effect is also interesting, peculiar to the rotating Cooper pairs.
\cite{Goryo00}

The most important point in this paper is that for the $p_x + i p_y$-wave case
one of the Cooper pairing electrons is coupled to a local spin
and the other has no direct coupling with the local spin.
As we will discuss later, the latter can interact with the local spin through the
superconducting order parameter.
We use the NRG method \cite{Wilson,Krishna,Sakai92}
to study this new type of two-channel Kondo effect.
As for the $s$-wave case, we discretize the bare conduction band logarithmically and
apply the Bogoliubov transformation to obtain a diagonal form of NRG Hamiltonian for
the superconductivity.
In the NRG Hamiltonian,
the two independent channels consisting of the $l = 0$ and $l = 1$ orbitals are coupled to
each other at the impurity.
The NRG result shows that the ground state is a spin doublet over all
the $T_{\rm K} / \Delta$ region and $T_{\rm K} / \Delta$ dependence of a bound state
expresses a competition between the Kondo effect and the superconducting energy gap.
The solution of the strong coupling limit ($T_{\rm K} / \Delta \rightarrow \infty$)
shows that the ground state is a spin doublet due to dynamical coupling between the $l = 0$
and $l = 1$ electrons,
while the local spin is almost quenched by the $l=0$ electrons.

This paper is organized as follows.
In Sec.~II we derive the NRG Hamiltonian to investigate the two-channel Kondo effect specific to
the $p_x + i p_y$ (or $d_{x^2 - y^2} + i d_{xy}$)-wave superconductivity.
Here the origin of the two channels is explained thoroughly.
In Sec.~III we show the NRG results in comparison with the $s$-wave case.
We also calculate a local spin susceptibility and discuss the Ising case.
It is very useful to understand the quantum effect of the magnetic impurity
in the gapped electron system.
This paper is closed with concluding remarks in Sec. IV.

\section{Formulation}
In this section we derive the NRG Hamiltonian
for the Kondo effect in
the $p_x + i p_y$-wave ($d_{x^2 - y^2} + i d_{xy}$-wave) superconductor.
Here we study a single magnetic impurity
at the center in two-dimensional superconducting systems.

\subsection{Model Hamiltonian}
Let us begin with the following Hamiltonian:
\begin{mathletters}
\begin{eqnarray}
&& H=H_{\rm kin} + H_\Delta + H_{\rm imp}, \\
&& H_{\rm kin} = \sum_{\bsk\sigma} \varepsilon_\bsk a_{\bsk\sigma}^\dagger a_{\bsk\sigma}, \\
&& H_\Delta = \sum_\bsk \left( \Delta_\bsk a_{\bsk\uparrow}^\dagger a_{-\bsk\downarrow}^\dagger
                             + {\rm H. c.} \right), \\
&& H_{\rm imp} = \frac{1}{2} \sum_{\bsk\bsk'\sigma\sigma'}
   \left( -J \bS \cdot \bsigma_{\sigma\sigma'} + V \delta_{\sigma,\sigma'} \right)
   a_{\bsk\sigma}^\dagger a_{\bsk'\sigma'}.
\end{eqnarray}
\label{eqn:H0}
\end{mathletters}\noindent
Here $\bk$ and $\sigma$ represent the wave vector and spin of the conduction electrons,
respectively.
$a_{\bsk\sigma}$ is the annihilation operator of the conduction electron,
$\epsilon_\bsk$ is the kinetic energy of the conduction electron,
and $\Delta_\bsk$ is the momentum dependent order parameter.
$\mbox{\boldmath$S$}$ expresses the $S=1/2$ spin operator for the magnetic impurity,
and $\mbox{\boldmath$\sigma$}$ is the Pauli matrix for the conduction electrons.
$J(<0)$ and $V$ are the antiferromagnetic and non-magnetic couplings, respectively.
Here we have assumed a short-range ($s$-wave) scattering impurity.
For the $p_x +i p_y$-wave pairing state,
the order parameter has a form of $\Delta_\bsk = \Delta e^{i\phi_\sk}$.
Here $\Delta$ is the order parameter which we assume a real number,
and $\phi_\sk$ is the angle of the wave vector measured from the $k_x$ axis.
Since the orbital angular momentum is a good quantum number for the $p_x +i p_y$-wave,
it is convenient to transform the wave vector representation
into the following two-dimensional polar coordinate representation:
\begin{equation}
a_{\bsk\sigma} = \sum_l (-i)^l |J_{l+1}(kR)| e^{i l \phi_\sk} a_{\sk l\sigma},
\label{eqn:plane}
\end{equation}
where $J_{l}$ is the $l$-th Bessel function,
and $R$ is the radius of the two-dimensional system.
$l$ is the $z$-component of the orbital angular momentum of the conduction electron.
Substituting Eq. (\ref{eqn:plane}) into Eq. (\ref{eqn:H0}), we obtain
\begin{mathletters}
\begin{eqnarray}
&& H = B \left( \bar{H}_{\rm kin} + \bar{H}_\Delta + \bar{H}_{\rm imp} \right), \\
&& \bar{H}_{\rm kin}
   = \int_{-1}^1 dk k \sum_{l\sigma} a_{\sk l \sigma}^\dagger a_{\sk l \sigma}, \\
&& \bar{H}_\Delta = \int_{-1}^1 dk \sum_l (-1)^{1-l}
   \left( i\bar{\Delta} a_{\sk l\uparrow}^\dagger a_{\sk,-l+1,\downarrow}^\dagger
   + {\rm H. c.} \right), \\
&& \bar{H}_{\rm imp} = \sum_{\sigma\sigma'}
   \left( -\bar{J} \bS \cdot \bsigma_{\sigma\sigma'} + \bar{V} \delta_{\sigma,\sigma'} \right)
   f_{00\sigma}^\dagger f_{00\sigma'}, \\
&& f_{00\sigma} = \frac{1}{\sqrt{2}} \int_{-1}^1 dk a_{\sk,l=0,\sigma},
\end{eqnarray}
\label{eqn:H1}
\end{mathletters}
\noindent
where $B=k_{\rm F}^3 R/(\pi m)=2k_{\rm F}R E_{\rm F}/\pi$,
$\bar{\Delta}=\Delta/(2E_{\rm F})$, $\bar{J}=J\rho$, $\bar{V}=V\rho$.
Here $E_{\rm F}=k_{\rm F}^2/(2m)$ is the Fermi energy,
and $\rho=mR^2/2=(k_{\rm F}R)^2/(4E_{\rm F})$ is the density of states at the Fermi energy
for the two-dimensional conduction band.
In Eq. (\ref{eqn:H1}) we have used the following asymptotic form of the Bessel function:
$J_l(kR) \simeq \sqrt{2/(\pi kR)}$ (for a large $kR$).
The wave number $k$ in Eq. (\ref{eqn:H1}) is normalized by $k_{\rm F}$.
The kinetic energy of the conduction electron is linearized near the Fermi energy.
For simplicity, we have used the same values
for both the superconducting cutoff energy and the band width $k_{\rm F}$,
since it is assured that this does not alter the results.
\cite{Sakai}
The fermion represented by the operator $f_{00\sigma}$ corresponds to a fermion localized
at the impurity.
The first subscript $0$ of $f_{00\sigma}$ represents the 0-th shell of the NRG Hamiltonian
which will be derived later (see Appendix).
The second zero means the $l = 0$ orbital.
Notice that the angular momentum of the $p_x +i p_y$-wave Cooper pair equals to one
(see $\bar{H}_{\rm \Delta}$),
and that the $l=0$ and $l=1$ orbitals are coupled.
Since the angular momentum is a good quantum number for the $p_x +i p_y$-wave,
both orbitals are decoupled from the other orbitals.
Due to the short-range ($s$-wave) scattering,
the electrons only in the $l=0$ orbital couple with the impurity.
The electrons in the $l=1$ orbital can also interact with the impurity
via the $p_x +i p_y$-wave order parameter.
It is sufficient to choose only the $l=0$ and $l=1$ orbitals in Eq. (\ref{eqn:H1})
to study the Kondo effect for the $p_x +i p_y$-wave state.
Therefore we can apply the NRG method to this Kondo problem
as we do to the two-channel Kondo problem.
\cite{Cragg,Pang}
We rewrite then $\bar{H}_{\rm kin}$ and $\bar{H}_{\rm \Delta}$ as
\begin{mathletters}
\begin{eqnarray}
&& \bar{H}_{\rm kin} = \int_{-1}^1 dk k \sum_\sigma
   \left( a_{\sk 0 \sigma}^\dagger a_{\sk 0 \sigma}
        + a_{\sk 1 \sigma}^\dagger a_{\sk 1 \sigma} \right), \\
&& \bar{H}_\Delta = \int_{-1}^1 dk
   \sum_\sigma \left( i\bar{\Delta} a_{\sk 1\sigma}^\dagger a_{\sk 0,-\sigma}^\dagger
                    + {\rm H. c.} \right).
\end{eqnarray}
\end{mathletters}
\noindent
We can obtain a similar ${\bar H}_\Delta$ for a $d_{x^2-y^2}+id_{xy}$-wave
if we replace $i\bar{\Delta}$ by $-\sigma\bar{\Delta}$ and $l=1$ by $l=2$,
since the order parameter for the $d_{x^2-y^2}+id_{xy}$-wave has a form of
$\Delta_\bsk=\Delta e^{i2\phi_\sk}$.

\subsection{Transformation of the Hamiltonian}
In our formulation we follow the procedure by Sakai {\it et al}.
\cite{Sakai}
and derive a NRG Hamiltonian.
There is another way to derive the Hamiltonian,
\cite{Satori}
which we describe in Appendix.
First we diagonalize the conduction electron part
by introducing the following Bogoliubov transformation:
\begin{mathletters}
\begin{eqnarray}
&& a_{\sk 0 \sigma} = u_\sk \alpha_{\sk,+,\sigma} - v_\sk^* \alpha_{\sk,-,-\sigma}^\dagger, \\
&& a_{\sk 1 \sigma} = u_\sk \alpha_{\sk,-,\sigma} + v_\sk^* \alpha_{\sk,+,-\sigma}^\dagger.
\end{eqnarray}
\end{mathletters}\noindent
Here $\alpha_{\sk,\tau=\pm,\sigma}$ is the annihilation operator
for the superconducting quasiparticle with $\sigma$ spin in $\tau$ channel.
The coefficients $u_\sk$ and $v_\sk$ are defined by
\begin{mathletters}
\begin{eqnarray}
&& u_\sk = \sqrt{ \frac{1}{2} (1+\frac{|k|}{E_\sk}) }, \\
&& v_\sk = i \sqrt{ \frac{1}{2} (1-\frac{|k|}{E_\sk}) } {\rm sgn}(k),
\end{eqnarray}
\end{mathletters}\noindent
where $E_\sk = (k^2 + \bar{\Delta}^2)^{1/2}$ is the energy of the quasiparticles.
The $p_x +i p_y$-wave Hamiltonian is then expressed
in terms of the superconducting quasiparticle as
\begin{eqnarray}
&& \bar{H}_{p_x +i p_y} = \bar{H}_{\rm kin} + \bar{H}_\Delta \cr
&& ~~~~~~~~~~= \int_{-1}^1 dk E_\sk {\rm sgn}(k) \sum_{\tau=\pm,\sigma}
         \alpha_{\sk\tau\sigma}^\dagger \alpha_{\sk\tau\sigma}.
\label{eqn:H-p}
\end{eqnarray}
For the later convenience,
we introduce the following transformations:
\begin{mathletters}
\begin{eqnarray}
&& \beta_{|k|,-\tau,-\sigma}^\dagger =  \alpha_{-|k|,\tau,\sigma}, \\
&& A_{\sk\tau\sigma} = \frac{1}{\sqrt{2}}
   \left( \alpha_{\sk\tau\sigma} - i\tau \beta_{\sk\tau\sigma} \right), \\
&& B_{\sk\tau\sigma} = \frac{1}{\sqrt{2}}
   \left( \alpha_{\sk\tau\sigma} + i\tau \beta_{\sk\tau\sigma} \right), \\
&& C_{+,\sigma}(k) = A_{\sk,+,\sigma}, \\
&& C_{+,\sigma}(-k) = -i B_{\sk,-,-\sigma}^\dagger, \\
&& C_{-,\sigma}(k) = i A_{\sk,-,-\sigma}^\dagger, \\
&& C_{-,\sigma}(-k) = B_{\sk,+,\sigma}.
\end{eqnarray}
\end{mathletters}\noindent
The Hamiltonian (\ref{eqn:H-p}) can be expressed by the above $C$ operators as
\begin{equation}
\bar{H}_{p_x +i p_y}
= \int_0^1 dk E_\sk \sum_{s=\pm,\tau,\sigma}
         \tau s C_{\tau\sigma}^\dagger(sk) C_{\tau\sigma}(sk).
\end{equation}

For the impurity part,
the operator $f_{00\sigma}$ is also expressed in terms of the superconducting quasiparticle as
\begin{mathletters}
\begin{eqnarray}
&& f_{00\sigma} = \frac{1}{\sqrt{2}} \int_0^1 dk
   \Bigl[ u_\sk \bigl( \alpha_{\sk+\sigma} + \beta_{\sk-,-\sigma}^\dagger \bigr) \cr
&&~~~~~~~~~~~~~~~~~~~~
        - v_\sk^* \bigl( \alpha_{\sk-,-\sigma}^\dagger - \beta_{\sk+,\sigma} \bigr)
   \Bigr] \cr
&& ~~~~~= \frac{1}{\sqrt{2}} \int_0^1 dk
   \Bigl[ w_{\sk+} \bigl( A_{\sk+\sigma} + i A_{\sk-,-\sigma}^\dagger \bigr) \cr
&&~~~~~~~~~~~~~~~~~~~~~
        + w_{\sk-} \bigl( B_{\sk+\sigma} - i B_{\sk-,-\sigma}^\dagger \bigr)
   \Bigr], \\
&& w_{\sk,\pm} = \sqrt{\frac{1}{2} \left( 1 \pm \frac{|\bar{\Delta}|}{E_\sk} \right)}.
\end{eqnarray}
\end{mathletters}\noindent
By introducing the following operators
\begin{mathletters}
\begin{eqnarray}
&& a_{\tau,\sigma} = N_+^{-1} \int_0^1 dk~w_{\sk +} A_{\sk,\tau,\sigma}, \\
&& b_{\tau,\sigma} = N_-^{-1} \int_0^1 dk~w_{\sk -} B_{\sk,\tau,\sigma}, \\
&& N_\pm = \left( \int_0^1 dk~w_{\sk \pm}^2 \right)^{1/2},
\end{eqnarray}
\end{mathletters}\noindent
$f_{00\sigma}$ is expressed as
\begin{eqnarray}
&&f_{00\sigma} = \frac{1}{\sqrt{2}}
   \Bigl[ N_+ a_{+,\sigma} - i N_- b_{-,-\sigma}^\dagger \cr
&&~~~~~~~~~~~~
        + i \bigl( N_+ a_{-,-\sigma}^\dagger - i N_- b_{+,\sigma}^\dagger \bigr)
   \Bigr] \cr
&& ~~~~~= \frac{1}{\sqrt{2}} \left( c_{0,+,\sigma} + c_{0,-,\sigma} \right),
\label{eqn:f0}
\end{eqnarray}
where $c_{0,\tau,\sigma}$ is defined by
\begin{mathletters}
\begin{eqnarray}
&& c_{0,+,\sigma} = N_+ a_{+,\sigma} - i N_- b_{-,-\sigma}^\dagger \cr
&&~~~~~~~=
   \int_0^1 dk \left( w_{\sk +} A_{\sk,+,\sigma}
                   -i w_{\sk -} B_{\sk,-,-\sigma}^\dagger \right) \cr
&&~~~~~~~=
   \int_0^1 dk \sum_{s=\pm} w_{\sk,s} C_{+,\sigma}(sk), \\
&& c_{0,-,\sigma} = i \left( N_+ a_{-,-\sigma}^\dagger - i N_- b_{+,\sigma} \right) \cr
&&~~~~~~~=
   \int_0^1 dk \left( w_{\sk +} i A_{\sk,-,-\sigma}^\dagger
                    + w_{\sk -} B_{\sk,+,\sigma} \right) \cr
&&~~~~~~~=
   \int_0^1 dk \sum_{s=\pm} w_{\sk,s} C_{-,\sigma}(sk).
\end{eqnarray}
\end{mathletters}\noindent
The exchange Hamiltonian
$\bar{H}_{\rm imp}$ is then written as
\begin{equation}
\bar{H}_{\rm imp} = \frac{1}{2} \sum_{\tau\tau'\sigma\sigma'}
   \left( -\bar{J} \bS \cdot \bsigma_{\sigma\sigma'} + \bar{V} \delta_{\sigma,\sigma'} \right)
   c_{0\tau\sigma}^\dagger c_{0\tau'\sigma'}.
\label{eqn:impurity}
\end{equation}
In Eq. (\ref{eqn:impurity}),
the subscript $\tau=\pm$ represents the two channels,
both of which consists of the $l=0$ and $l=1$ orbitals.
While the impurity interacts with only the $l=0$ electron,
Eq. (\ref{eqn:impurity}) contains the channel-flip ($\sum_{\tau\tau'}$) terms.
This is because the operator $f_{00\sigma}$ is expressed as Eq. (\ref{eqn:f0}).
The two-channel property in Eq. (\ref{eqn:impurity})
originates from the angular momentum of the $p_x +i p_y$-wave Cooper pair.
In order to treat the Kondo problem in the superconducting state,
first we have diagonalized the Hamiltonian of the conduction electron part.
The diagonalized representation ($\tau$ representation) in this Hamiltonian
is given by the linear combination of the $l=0$ and $l=1$ orbitals,
since the $p_x +i p_y$-wave mixes the two orbitals.
It is natural that the impurity part of the Hamiltonian contains the channel-flip part
in the $\tau$ representation.

In contrast to the present $p_x +i p_y$-wave case,
the standard two-channel Kondo model
was derived from the internal degrees of freedom of the impurity atom.
\cite{Noziers,Cox}
In our model,
the impurity does not have such internal degrees of freedom except for the $S=1/2$ spin.
However, we can obtain the two-channel property
due to the orbital effect of the Cooper pairs.
Therefore the present two-channel property for the $p_x +i p_y$-wave
can provide new Kondo physics peculiar to the unconventional superconductors.

\subsection{Discretization}
Following Wilson's procedure,\cite{Wilson}
we discretize the conduction band with a logarithmic mesh in the $k$ space.
The Hamiltonian for the discretized conduction band is written as
\begin{mathletters}
\begin{eqnarray}
&& \bar{H}_{p_x +i p_y} = \sum_{m=0}^\infty \sum_{\tau \sigma s}
   \tau s E_m C_{\tau\sigma}^\dagger(sm) C_{\tau\sigma}(sm), \\
&& E_m = \sqrt{k_m^2 + \bar{\Delta}^2}, \\
&& k_m = (1+\Lambda^{-1})\Lambda^{-m}/2,
\end{eqnarray}
\label{eqn:discre}
\end{mathletters}\noindent
where $\Lambda >1$ is the discretization parameter.
The above $C$ operators satisfies the Fermion commutation relation:
\begin{equation}
\left\{ C_{\tau\sigma}(sm), C_{\tau'\sigma'}^\dagger(s'm') \right\}
  = \delta_{\tau,\tau'} \delta_{\sigma,\sigma'} \delta_{s,s'} \delta_{m,m'}.
\end{equation}
The fermion operator for the localized orbit (the 0-th shell orbit) $c_{0\tau\sigma}$ is
also expressed with the $C$ operators as
\begin{mathletters}
\begin{eqnarray}
&& c_{0\tau\sigma}
   = (1-\Lambda^{-1})^{-1/2} \sum_{sm} w_{ms} \Lambda^{-m/2} C_{\tau\sigma}(sm), \\
&& w_{ms} = \sqrt{(1 + s \frac{\bar{\Delta}}{E_m})/2}.
\end{eqnarray}
\end{mathletters}\noindent
We define $c_{n\tau\sigma}$ for the n-th shell orbit and satisfies the following commutation
relation:
\begin{eqnarray}
\left[ c_{n\tau\sigma}, \bar{H}_{p_x +i p_y} \right]
  &=& t_{n-1} c_{n-1,\tau\sigma} + \tau \eta_n c_{n\tau\sigma} \cr
  &+& t_n c_{n+1,\tau\sigma}.
\end{eqnarray}
For $n=0$ we have
\begin{eqnarray}
\left[ c_{0\tau\sigma}, \bar{H}_{p_x +i p_y} \right]
   &=& (1-\Lambda^{-1})^{1/2} \cr
&& \times\sum_{sm} w_{ms} \Lambda^{-m/2} \tau s E_m C_{\tau\sigma}(sm) \cr
   &=& \tau \eta_0 c_{0\tau\sigma} + t_0 c_{1\tau\sigma},
\end{eqnarray}
where $\eta_0$, $t_0$, and $c_{1\tau\sigma}$ are defined by
\begin{mathletters}
\begin{eqnarray}
&& \eta_0 = (1-\Lambda^{-1}) \sum_{sm} w_{ms}^2 \Lambda^{-m} s E_m, \\
&& t_0 = \left[ (1-\Lambda^{-1})\sum_{sm} w_{sm}^2 \Lambda^{-m}(sE_m - \eta_0)^2 \right]^{1/2}, \\
&& c_{1\tau\sigma} = \tau \frac{(1-\Lambda^{-1})^{1/2}}{t_0} \cr
&&~~~~~~\times \sum_{sm} w_{sm} \Lambda^{-m/2} (sE_m -\eta_0) C_{\tau\sigma}(sm).
\end{eqnarray}
\end{mathletters}\noindent
Repeating this procedure, we obtain the total Hamiltonian which has the following hopping
type with staggered potentials:
\cite{Sakai}
\begin{mathletters}
\begin{eqnarray}
&& \bar{H} = \bar{H}_{p_x +i p_y} + \bar{H}_{\rm imp}, \\
&& \bar{H}_{p_x +i p_y} 
   = \sum_{n\tau\sigma} \Bigl[ \tau \eta_n c_{n\tau\sigma}^\dagger c_{n\tau\sigma} \cr
&&~~~~~~~~~~~ + t_n \left( c_{n+1,\tau\sigma}^\dagger c_{n\tau\sigma}
                   + c_{n,\tau\sigma}^\dagger c_{n+1,\tau\sigma} \right)
                        \Bigr], \\
&& \bar{H}_{\rm imp} = \frac{1}{2} \sum_{\tau\tau'\sigma\sigma'}
   \left( -\bar{J} \bS \cdot \bsigma_{\sigma\sigma'} + \bar{V} \delta_{\sigma,\sigma'} \right)
   c_{0\tau\sigma}^\dagger c_{0\tau'\sigma'}.
\end{eqnarray}
\label{eqn:discre2}
\end{mathletters}\noindent
Here $\eta_n$ and $t_n$ are given by
\begin{mathletters}
\begin{eqnarray}
&& \eta_n = (-1)^n \bar{\Delta}, \\
&& t_n = \frac{(1 + \Lambda^{-1})}{2} \varepsilon_n \Lambda^{-n/2}, \\
&& \varepsilon_n = \left[1-\Lambda^{-(n+1)}\right] \left[1-\Lambda^{-(2n+1)}\right]^{-1/2} \cr
&&~~~
\times \left[1-\Lambda^{-(2n+3)}\right]^{-1/2}.
\label{eqn:en}
\end{eqnarray}
\label{eqn:discritized}
\end{mathletters}\noindent
In the practical calculation, we use the following scaled Hamiltonian
in a recursion relation given by
\begin{mathletters}
\begin{eqnarray}
&&H_{N+1} = \Lambda^{1/2}H_N + \sum_{\tau\sigma}
  \Bigl[
    \varepsilon_N ( c_{N+1,\tau\sigma}^\dagger c_{N\tau\sigma} + {\rm H.c.} ) \cr
&&~~~~~~~~~~~~~~~
     +(-1)^N\Lambda^{N/2}\tau\tilde{\Delta}
        c_{N+1,\tau\sigma}^\dagger c_{N+1,\tau\sigma}
  \Bigr], \\
&&H_0 =
  \Bigl[ \frac{1}{2}
    \sum_{\tau\tau'\sigma\sigma'}
      ( - \tilde{J} \mbox{\boldmath$S$} \cdot \mbox{\boldmath$\sigma$}_{\sigma\sigma'}
        + \tilde{V} \delta_{\sigma,\sigma'} )
        c_{0\tau\sigma}^\dagger c_{0\tau'\sigma'} \cr
&&~~~~~~~~~~~~~~~ -\sum_{\tau\sigma} \tau\tilde{\Delta} c_{0\tau\sigma}^\dagger c_{0\tau\sigma}
  \Bigr] \Lambda^{-1/2}.
\end{eqnarray}
\label{eqn:NRG}
\end{mathletters}\noindent
Here the parameters are defined by
\begin{eqnarray}
&&\tilde{\Delta} = \frac{\bar{\Delta}}{(1+\Lambda^{-1})/2}, \cr
&&\tilde{J} = \frac{\bar{J}}{(1+\Lambda^{-1})/2}, \cr
&&\tilde{V} = \frac{\bar{V}}{(1+\Lambda^{-1})/2}.
\end{eqnarray}
The Hamiltonian (\ref{eqn:discre2}) is related to $H_N$ and is given by
\begin{equation}
\bar{H} = \frac{1+\Lambda^{-1}}{2} \lim_{N\rightarrow \infty} \Lambda^{-(N-1)/2} H_N.
\end{equation}
In Eq. (\ref{eqn:NRG}),
$c_{N\tau\sigma}$ is an operator of the NRG fermion quasiparticle in the $N$-th shell.
The subscript $\tau=\pm$ represents the two channels
which are constructed by the $l=0$ and $l=1$ orbitals
as discussed in the last part of the previous subsection.
We can obtain the same NRG Hamiltonian for a $d_{x-2-y^2} +i d_{xy}$-wave.
In order to check reliability of numerical results, we compare different data
by changing cutoff of the eigenstates and the discretization parameter $\Lambda$.
In the NRG results we show throughout this paper,
we keep the lowest-lying $\sim 500$ states at each renormalization step and
take $\Lambda=3$.
For calculating a local spin susceptibility, we keep $\sim 1200$ states
to reduce the cutoff dependence.

\section{Kondo Effect in $p$-wave Superconductor}
In this section we present our results for the Kondo effect
in the $p_x +i p_y$-wave superconducting state.
We show numerical results
and then discuss the strong coupling limit to see
how the orbital dynamics of the Cooper pairs affects the Kondo singlet.

\subsection{Numerical results}

First we discuss a $\tilde{\Delta}=0$ case
where the only $l=0$ electrons couple with the local spin
and the $l=1$ electrons are free from the impurity.
The NRG result shows that the Kondo singlet is realized by the relevant $l=0$ electrons
at the stable fixed point.
This fixed point is of strong coupling type
($|\tilde{J}| \rightarrow \infty$)
and the fixed-point NRG energy levels is reproduced by truncating
the $N=0$ site (impurity site) from the $l=0$ chain of the conduction electrons.
\cite{Wilson}
On the other hand,
the fixed point for $l=1$ is of free-electron type ($\tilde{J} =0$).
One of the two chains produces a zero energy state
in one-particle energy levels at the fixed point.
We can fill at most two particles (spin up and down) into the zero energy state
and we have quartet ground state for $\tilde{\Delta}=0$.

Next we discuss a $\tilde{J}=\tilde{V}=0$ case with finite $\tilde{\Delta}$.
In this case,
the NRG Hamiltonian (\ref{eqn:NRG}) consists of the two independent $\tau=\pm$ channels
related to the superconducting state.
The Hamiltonian of each channel is same as that of the free $s$-wave superconductivity.
\cite{Satori}
At the odd renormalization steps,
one-particle energy level for the free $s$-wave is particle-hole symmetric.
At the even steps,
there is an additional one-particle energy level which is hole-like (particle-like)
for a positive (negative) $\tilde{\Delta}$.
In the present $p_x +i p_y$-wave case,
one particle energy level is symmetric at all the renormalization steps,
since the signs of $\tilde{\Delta}$ for the $\tau=\pm$ channels are different each other.
In one-particle energy levels,
there is no zero energy state due to the finite $\tilde{\Delta}$.
The ground state is obtained by filling particles into the energy levels
and we have a single ground state.

\begin{figure}[t]
\begin{center}
\epsfxsize=7cm
\epsfbox{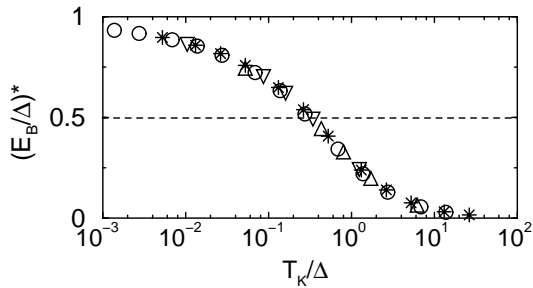}
\end{center}
\caption{
$T_{\rm K}/\Delta$ dependence of the bound state energy $(E_{\rm B}/\Delta)^*$
for $p_x +i p_y$-wave.
The ground state is always a spin doublet
and the bound state level is measured from the ground state for each $T_{\rm K}/\Delta$.
The first excited state is a particle-hole doublet with no spin for $\tilde{V} = 0$.
The circle and star represent the result with fixed $T_{\rm K}$
for $T_{\rm K}=9.20\times 10^{-5}$ and $1.76\times 10^{-3}$, respectively.
The triangle-up and triangle-down are the results for fixed $\tilde{\Delta}=0.001$ and $0.005$,
respectively.
Here $\tilde{\Delta}=1.5\Delta$.
For both $T_{\rm K}$ and $\Delta$,
a unit of the energy is the band width $2E_{\rm F}$.
}
\label{fig:1}
\end{figure}
Let us turn to the Kondo problem in the $p_x +i p_y$-wave state.
In this case both $\tilde{\Delta}$ and $\tilde{J}$ are finite.
Since $\tilde{J}$ grows up as $\Lambda^{N/2}$ with increasing $N$,
it is relevant in the Kondo effect.
In addition,
$\tilde{\Delta}$ is also relevant since it grows as in Eq. (\ref{eqn:NRG}).
Therefore we can expect a competition between the Kondo effect and the superconductivity.
\cite{Satori}
As the renormalization step increases,
the excited state energies increase with $\tilde{\Delta}_N \sim \Lambda^{N/2}$,
which is the energy gap at the $N$-th renormalization step.
Since the magnetic impurity can destroy the superconductivity,
we have bound states below $\tilde{\Delta}_N$.
The ratio of the bound state energy and $\tilde{\Delta}_N$ approaches a constant value
with the increase of the renormalization step.
We express the convergent value of the ratio by $(E_{\rm B}/\Delta)^*$.
Figure \ref{fig:1} shows the $T_{\rm K}/\Delta$ dependence of the first excited state
(bound state) for $\tilde{V} = 0$,
where the Kondo temperature is defined by
\begin{equation}
T_{\rm K}=\sqrt{|\bar{J}|}{\rm exp}(-1/|\bar{J}|).
\end{equation}
Here, a unit of the energy is taken to be the band width $2E_{\rm F}$.
We find that the bound state energy level is scaled by $T_{\rm K}/\Delta$
as in the $s$-wave case.
In a small $T_{\rm K}/\Delta$ region,
the ground state is a spin doublet as expected,
and the first excited state is a particle-hole doublet with no spin.
With the increase of $T_{\rm K}/\Delta$,
the energy level of the bound state (the first excited state) decreases (see Fig. \ref{fig:1}).
As $T_{\rm K}/\Delta \rightarrow \infty$,
it approaches the ground state energy and the ground state is four-fold degenerate.
The result for $T_{\rm K}/\Delta \rightarrow \infty$
is connected smoothly to that for $\Delta = 0$.

The most important point in Fig. \ref{fig:1} is
that the ground state is always a spin doublet.
This means that the spin of the ground state survives even in the strong coupling limit
for the $p_x +i p_y$-wave superconductivity.
In the $s$-wave case,
the Kondo singlet is realized in the large $T_{\rm K}/\Delta$ region
\cite{Satori},
resulting in a spin singlet ground state.
Thus the Kondo effect for the $p_x +i p_y$-wave is quite different
from that for the conventional $s$-wave.

While the ground state is always a spin doublet for the $p_x +i p_y$-wave,
we find that the $T_{\rm K}/\Delta$ dependence of the bound state energy level
is very similar to that of the $s$-wave.
For the $s$-wave,
the Kondo singlet is realized in $T_{\rm K}/\Delta > 0.3$ region
against the superconducting energy gap.
\cite{Satori}
In Fig. \ref{fig:1},
the curvature of the $(E_{\rm B}/\Delta)^*$ curve found around $(E_{\rm B}/\Delta)^*=0.5$
is very similar to that of the $s$-wave case
found with the interchange of the ground state.
This represents a crossover from the $\Delta$ dominant region to the $T_{\rm K}$ dominant region.
It occurs around $T_{\rm K}/\Delta=0.3$ as in the $s$-wave case.
Thus the Kondo effect can overcome the energy gap also in the $p_x +i p_y$-wave case,
although the ground state is still a spin doublet.
This means that the competition between the Kondo effect and the energy gap
is characterized by such $T_{\rm K}/\Delta$ dependence.
All the above results hold for the $d_{x^2-y^2} +i d_{xy}$-wave as well,
since we can obtain the same NRG Hamiltonian.

Finally,
we mention the effect of the potential scattering $\tilde{V}$ in Eq. (\ref{eqn:NRG}).
It is well known that a single non-magnetic impurity destroys the unconventional superconductivity.
It can generate a bound state locally around the impurity.
If we treat the magnetic impurity realistically,
we have to start from the Anderson model.
Since the generalized Kondo model consists of both magnetic and non-magnetic interactions,
we have to take both couplings into account.
Since the effect of $\tilde{V}$ breaks the particle-hole symmetry,
it lifts the degeneracy of the first excited particle-hole doublet.
However, the ground state is still a spin doublet
and $\tilde{V}$ is irrelevant even if it is large enough.

\subsection{Strong coupling limit}
\begin{figure}[t]
\input{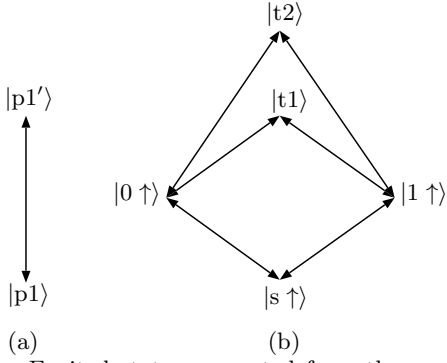}
\caption{
Excited states generated from the unperturbed ground ones by the perturbation of $\tilde{\Delta}$.
(a) Particle-hole doublet case.
(b) Spin doublet case.
}
\label{fig:2}
\end{figure}

Here we study the mechanism of the spin doublet ground state for the $p_x +i p_y$-wave.
Let us discuss the strong coupling limit ($|\tilde{J}|\rightarrow \infty$) case.
The Hamiltonian $H_0$ in Eq. (\ref{eqn:NRG}) can exhibit the solution in this limit.
Returning the channel ($\tau$) representation to the angular momentum ($l$) one,
we can rewrite the $H_0$ for $\tilde{V} = 0$ as
\begin{eqnarray}
&& H_0 = H_{0J} + H_{0\Delta}, \cr
&& H_{0J} = -\tilde{J} \sum_{\sigma\sigma'}
  \mbox{\boldmath$S$} \cdot \mbox{\boldmath$\sigma$}_{\sigma\sigma'}
  g_{0\sigma}^\dagger g_{0\sigma'}, \cr
&& H_{0\Delta} = - \tilde{\Delta} \sum_\sigma
  ( g_{0\sigma}^\dagger g_{1\sigma} + g_{1\sigma}^\dagger g_{0\sigma} ),
\label{eqn:limit-p}
\end{eqnarray}
where the following transformation has been used:
\begin{equation}
c_{N=0,\tau,\sigma}=\frac{1}{\sqrt{2}} \left( g_{l=0,\sigma} +\tau g_{l=1,\sigma} \right).
\end{equation}
Here $g_{l = 0,\sigma}$ and $g_{l = 1,\sigma}$ correspond to $f_{00\sigma}$ and
$f_{01\sigma}^{\dag}$, respectively.
Notice that the order parameter corresponds to the hopping parameter
between the $l=0$ and $l=1$ local orbital sites,
since the $p_x +i p_y$-wave Cooper pair consists of the $l=0$ and $l=1$ particles.
When $|\tilde{J}| \rightarrow \infty$,
the $l=0$ particle couples with the local spin strongly to form a spin singlet $|s \rangle$:
\begin{equation}
|{\rm s} \rangle = \frac{1}{\sqrt{2}} \left( g_{0\uparrow}^\dagger |\downarrow \rangle
                                            -g_{0\downarrow}^\dagger |\uparrow \rangle \right),
\end{equation}
where $|\uparrow \rangle$ and $|\downarrow \rangle$ represent the local spin states.
Since the hopping is forbidden,
the ground states are 4-fold degenerate for $|J|\rightarrow\infty$:
\begin{eqnarray}
&& |{\rm s}\uparrow \rangle = g_{1\uparrow}^\dagger |{\rm s} \rangle, \cr
&& |{\rm s}\downarrow \rangle = g_{1\downarrow}^\dagger |{\rm s} \rangle, \cr
&& |{\rm p1} \rangle = |s \rangle, \cr
&& |{\rm p2} \rangle = g_{1\uparrow}^\dagger g_{1\downarrow}^\dagger |s \rangle,
\label{eqn:ground-p}
\end{eqnarray}
where $|{\rm s}\sigma\rangle$ ($\sigma=\uparrow,\downarrow$) and $|{\rm p}i\rangle$ ($i=1,2$)
are the spin doublet and particle-hole doublet, respectively.
At a finite $|\tilde{J}|$, $\tilde{\Delta}$ in Eq. (\ref{eqn:limit-p}) lifts the degeneracy.
The perturbation (the $\tilde{\Delta}$ term) generates excited states from the ground states
in Eq. (\ref{eqn:ground-p}).
We show the connection among these states in Fig. \ref{fig:2}.
One of the particle-hole doublet states is connected to a single excited state.
For $|{\rm p1}\rangle$, the excited state is given by [see Fig. \ref{fig:2} (a)]:
\begin{equation}
 |{\rm p1}'\rangle = \frac{1}{\sqrt{2}}
  \left( g_{1\uparrow}^\dagger |\downarrow \rangle
        -g_{1\downarrow}^\dagger |\uparrow \rangle
  \right).
\end{equation}
The similar argument holds for $|{\rm p2}\rangle$.
We can obtain the following eigenvalue for the particle-hole doublet up to the fourth order of the
perturbation:
\begin{equation}
E_{S=0} =
  \left[ - \frac{3}{2}
         - \frac{2}{3} \left( \frac{\tilde{\Delta}}{|\tilde{J}|} \right)^2
         + \frac{8}{27} \left( \frac{\tilde{\Delta}}{|\tilde{J}|} \right)^4
\right] |\tilde{J}|.
\end{equation}

For one of the spin doublet states,
it is connected to four excited states by the perturbation up to the fourth order.
For $|{\rm s}\uparrow \rangle$, these excited states are given as follows
[see Fig.~\ref{fig:2} (b)]:
\begin{eqnarray}
&& |0 \uparrow \rangle = g_{0\uparrow}^\dagger g_{0\downarrow}^\dagger |\uparrow \rangle, \cr
&& |1 \uparrow \rangle = g_{1\uparrow}^\dagger g_{1\downarrow}^\dagger |\uparrow \rangle, \cr
&& |{\rm t1} \rangle = g_{1\downarrow}^\dagger g_{0\uparrow}^\dagger |\uparrow \rangle, \cr
&& |{\rm t2} \rangle = \frac{1}{\sqrt{2}} g_{1\uparrow}^\dagger
  \left[ g_{0\uparrow}^\dagger |\downarrow \rangle
       + g_{0\downarrow}^\dagger |\uparrow \rangle
  \right],
\end{eqnarray}
and the similar connection is obtained for $|{\rm s}\downarrow \rangle$ as well.
The energy eigenvalue can be obtained from the following equation:
\begin{eqnarray}
&& e(e-1/2)\left[ e^3 + e^2 - (4x^2 +3/4 )e - 4x^2 \right] = 0, \cr
&& e=E/|\tilde{J}|,~~~~~~x=|\tilde{\Delta}|/|\tilde{J}|.
\end{eqnarray}
After expanding the lowest $e$ in $x$ up to the fourth order,
we obtain the energy for the spin doublet as
\begin{equation}
E_{S=1/2} =
  \left[ - \frac{3}{2}
         - \frac{2}{3} \left( \frac{\tilde{\Delta}}{|\tilde{J}|} \right)^2
         - \frac{10}{27} \left( \frac{\tilde{\Delta}}{|\tilde{J}|} \right)^4
\right] |\tilde{J}|.
\end{equation}
The energy difference between the two doublets is given by
\begin{equation}
E_{S=0} - E_{S=1/2} = \frac{2}{3} |\tilde{J}|
  \left( \frac{\tilde{\Delta}}{\tilde{J}} \right)^4.
\end{equation}
Thus, the order parameter $\tilde{\Delta}$ stabilizes the spin doublet
even if it is very small.

In the $s$-wave case, the superconducting part of $H_0$ is given by
\begin{equation}
H_{0\Delta} = - \tilde{\Delta} \left( g_{0\uparrow}^\dagger g_{0\uparrow}
 + g_{0\downarrow}^\dagger g_{0\downarrow} -1 \right).
\end{equation}
We notice that it is not given by hopping between the sites but
is given by a potential on the $l=0$ site.
This is because the $s$-wave Cooper pair is formed by only the $l=0$ quasiparticles.
In the $|\tilde{J}|\rightarrow\infty$ limit, we have a Kondo singlet ground state as expected.

\begin{figure}[t]
\begin{center}
\input{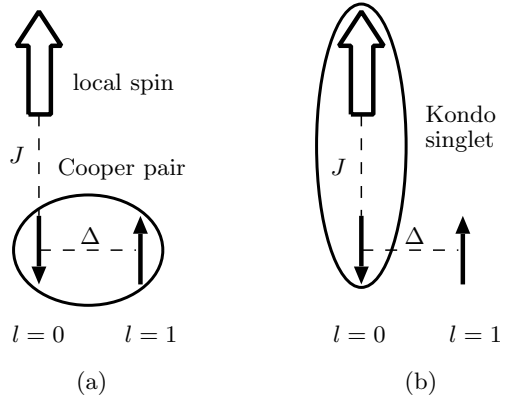}
\end{center}
\caption{
Schematic picture of the ground state wavefunction.
(a) Weak coupling limit ($T_{\rm K}/\Delta \rightarrow 0$).
(b) Strong coupling limit ($T_{\rm K}/\Delta \rightarrow \infty$).
}
\label{fig:3}
\end{figure}
Comparing the $p_x +i p_y$-wave result with the $s$-wave one,
we conclude that the spin of the ground state is realized
by the orbital dynamics of the $p_x +i p_y$-wave Cooper pairs.
This point is illustrated by Fig. \ref{fig:3}.
In a weak coupling region,
the spin of the ground state is mainly produced by the local spin itself
[see Fig. \ref{fig:3}(a)].
In the strong coupling limit, however,
a Kondo singlet is formed by the $l=0$ conduction electrons.
Therefore the local spin is almost quenched by the $l=0$ electrons.
In this limit, one of the paired electrons having $l=1$ angular momentum
is connected weakly with the Kondo singlet to gain the superconducting condensation energy.
The latter electrons generate the ground state spin as shown in Fig. \ref{fig:3}(b).

\subsection{Impurity susceptibility}
\begin{figure}[t]
\begin{center}
\epsfxsize=7cm
\epsfbox{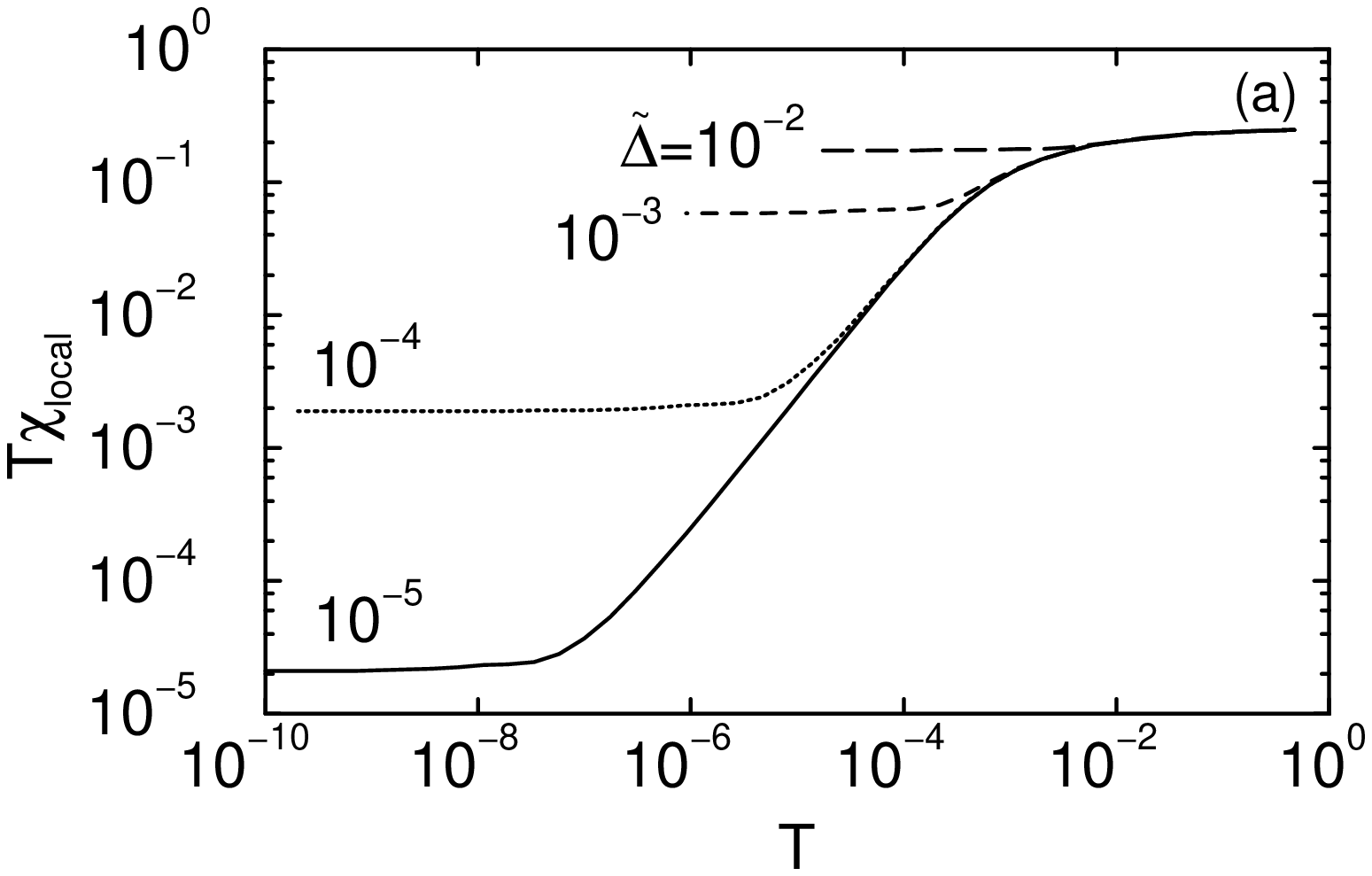}
\epsfxsize=7cm
\epsfbox{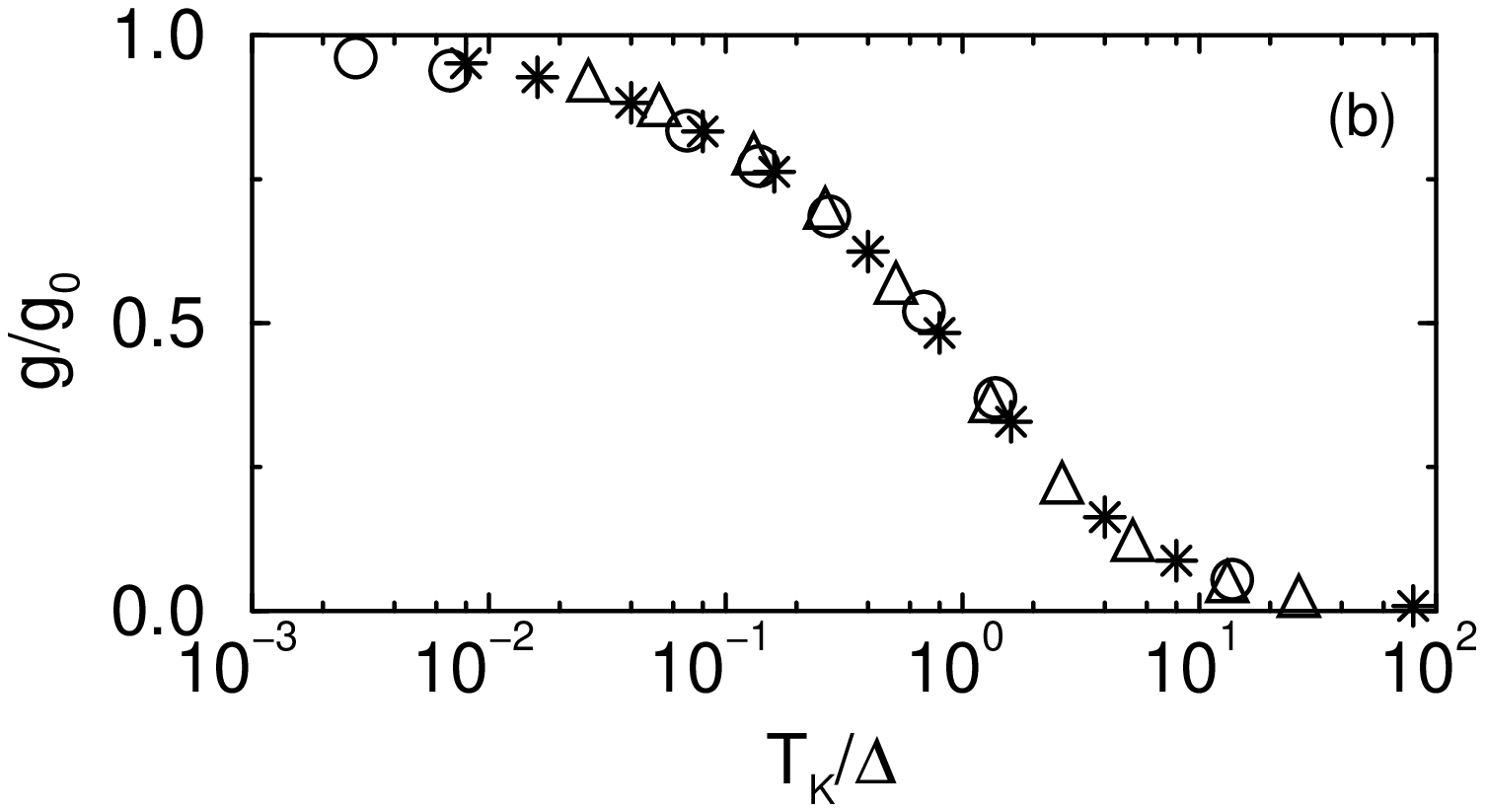}
\end{center}
\caption{
(a) Temperature dependence of $T \chi_{\rm local}$ for various $\tilde{\Delta}$.
Here $\tilde{\Delta}=1.5\Delta$.
The exchange coupling is fixed as $\tilde{J}=-0.25$ ($T_{\rm K}=5.34\times 10^{-4}$).
The unit is $(g_0\mu_{\rm B})^2$, where $g_0$ is a bare $g$-factor of the local spin.
(b) $T_{\rm K}/\Delta$ dependence of the effective $g$-factor.
The circle, star, and triangle-up represent the results
with fixed $T_{\rm K}=9.20\times 10^{-5}$, $5.34\times 10^{-4}$,
and $1.76\times 10^{-3}$, respectively.
For both $T_{\rm K}$ and $\Delta$,
a unit of the energy is the band width $2E_{\rm F}$.
}
\label{fig:4}
\end{figure}
Next we see how the spin of the
magnetic impurity is screened by the Kondo effect.  For this purpose
we calculate a local spin susceptibility which is defined by
\begin{mathletters}
\begin{eqnarray}
\chi_{\rm local}(T) &=& g\mu_{\rm B}\frac{\langle S_z \rangle}{h}|_{h\rightarrow 0}, \\
\langle S_z \rangle &=& \frac{ {\rm Tr} \left[S_z {\rm exp}(-\bar{\beta} H_N)\right]}
                  { {\rm Tr} \left[{\rm exp}(-\bar{\beta} H_N)\right]}.
\end{eqnarray}
\end{mathletters}\noindent
Here $S_z$ is the local spin operator,
and $\bar{\beta}\sim 1$ is taken in the calculation.
The temperature $T$ is given by
\cite{Wilson,Krishna,Sakai92}
\begin{equation}
T = \frac{1}{2}(1+\Lambda^{-1})\Lambda^{-(N-1)/2}/\bar{\beta}.
\end{equation}
A small magnetic field $h$ is applied only at the local spin site.
The impurity part of the Hamiltonian $H_0$ in Eq. (\ref{eqn:NRG}b) is replaced as
\begin{equation}
H_0 \rightarrow H_0 - g\mu_{\rm B} S_z h \Lambda^{-1/2}.
\end{equation}
Therefore $\chi_{\rm local}$ represents magnetic response of the local spin.
In Fig. \ref{fig:4}(a) we show the temperature dependence of $\chi_{\rm local}$
for various $\Delta$ with a fixed exchange coupling $\tilde{J}$.
We can see that $T \chi_{\rm local}$ decreases monotonically
as we decrease temperature (increase the renormalization step) for each $\Delta$,
and that it becomes constant at low temperatures,
indicating the Curie law.
However, the effective $g$-factor of the local spin depends on both $\tilde{J}$ and $\Delta$
shown in Fig. \ref{fig:4}(b).
We can see that $g$ is scaled by $T_{\rm K}/\Delta$
as in the case of the energy of the first excited state [see Fig. \ref{fig:1}].
This result indicates that the local spin is actually screened by the Kondo effect,
although the ground state is a spin doublet.

\begin{figure}[t]
\begin{center}
\epsfxsize=7cm
\epsfbox{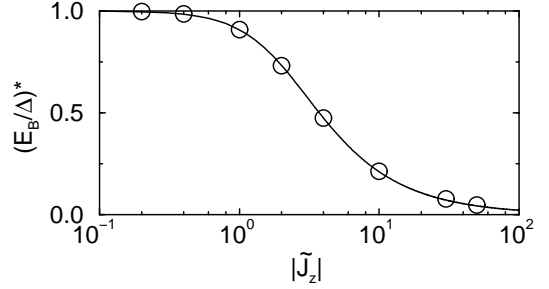}
\end{center}
\caption{
Bound state energy calculated by the NRG method for the Ising case.
It is normalized by $\Delta$.
The solid line is the analytic solution given by
$E_{\rm B}/\Delta = [1+(\alpha \tilde{J}_z)^2]^{-1/2}$,
where $\tilde{J}_z=1.5\bar{J}_z$, and $\alpha$ is a constant.
We have used $\alpha=0.465$ to fit the NRG result.
}
\label{fig:5}
\end{figure}
In order to see the quantum effect (Kondo effect) of the local spin,
we compare the result of the bound state energy for a Ising (classical) spin
[see Fig. \ref{fig:5}].
The bound state energy position is given by
\begin{mathletters}
\begin{eqnarray}
E_{\rm B} &=& \frac{\Delta}{\sqrt{1+\gamma^2}}, \\
\gamma &=& \frac{\pi N_0}{4 \Omega} |J_z|,
\end{eqnarray}
\end{mathletters}
which is essentially same as in the non-magnetic impurity case.
\cite{Okuno}
Here $N_0$ is the density of state at the Fermi energy in the normal state.
$\Omega$ is the volume of the system.
The bound state energy decreases as we increase the Ising coupling $|\tilde{J}_z|$.
This point is similar to the result of the quantum spin case shown in Fig. \ref{fig:1}.
However, the bound state energy is independent of $\Delta$ for the Ising spin,
while it is a function of $T_{\rm K}/\Delta$ for the quantum spin.
In addition, we have to use an extremely larger $|\tilde{J}_z|$ value
($|\tilde{J}_z| \gg 1$) for the Ising spin to obtain a small $(E_{\rm B} / \Delta)^*$.
This indicates that the spin-flip effect (quantum effect) is essential to our result,
and that the Kondo effect actually occurs in the $p_x +i p_y$-wave superconductor.

\section{Conclusion and Discussion}
We have studied the novel Kondo effect
due to the orbital effect of the Cooper pairs of unconventional superconductors.
In particular,
we have focused on the $p_x +i p_y$-wave ($d_{x^2-y^2}+id_{xy}$-wave) pairing state.
Regarding the magnetic impurity,
we have assumed an $s$-wave (short-range type) impurity scattering,
which is the simplest but is primarily important in real materials.
The characteristic point of the unconventional superconductor is
that the Cooer pair has orbital degrees of freedom.
Therefore the superconducting quasiparticles consist of electrons with various angular momenta.
The object of this paper is to clarify
how the orbital dynamics of the Cooper pairs affects the Kondo effect
in unconventional superconductors.
The followings are our main results.

\noindent
(1) $p_x +i p_y$-wave \par
In the superconducting state,
the low-energy excitations are moved out of the superconducting energy gap.
The $p_x +i p_y$-wave superconductor is a fully gapped system.
Unlike the $s$-wave superconductor,
the $p_x +i p_y$-wave Cooper pairs have the orbital degrees of freedom.
The local spin couples with only $l=0$ electrons
which are paired with the $l=1$ ones.
However, the $l=1$ electrons can interact with the local spin
via the superconducting order parameter.
We note that there is no direct coupling between the local spin and the $l=1$ electrons.
In contrast to the two-channel Kondo effect derived from the atomic structure of the impurity,
\cite{Noziers}
the present two-channel property is the consequence
of the internal degrees of freedom of the conduction electron system.
Our NRG result shows that
the ground state is a spin doublet over all the $T_{\rm K}/\Delta$ region.
This is completely different from the result of the $s$-wave superconductor
in which the Kondo singlet is realized in a large $T_{\rm K}/\Delta$ region.
We also find that the first excited energies are scaled by $T_{\rm K}/\Delta$.
In the strong coupling limit ($T_{\rm K}/\Delta \rightarrow \infty$),
it is found that the effective $g$-factor of the local spin
is strongly reduced by the Kondo effect,
while the $p_x + i p_y$-wave Cooper pair couples with the Kondo singlet,
generating the spin of the ground state.
Thus the orbital effect of the Cooper pair is important
in the Kondo effect we have discussed here.

\noindent
(2) Origins of the multichannel property \par
Generally, there are two origins of the multichannel property in the Kondo effect.
The first one is the internal degrees of freedom of the impurity atom
as pointed out by Nozi$\grave{\rm e}$res and Blandin.
\cite{Noziers}
In this case, the channels of the conduction electron system is controlled
by the atomic structure of the impurity atom.

The second one is related to the conduction electron systems.
In this paper we have investigated the Kondo effect in the unconventional superconductors.
As shown in the $p_x +i p_y$-wave case,
the multichannel property comes from the angular momenta of the Cooper pairs.
The independent channels of the superconducting quasiparticles
couple with each other at the impurity
even if there is no internal degrees of freedom in the impurity atom (just a local spin).
This is because the symmetry of the conduction electron system is different from
that of the impurity scattering.
As a different variety besides the superconducting systems,
we can introduce strong momentum dependence into the bare conduction band.
For example,
a singular pseudo-gap in the vicinity of the Fermi energy produces unusual Kondo behavior.
\cite{Gonzalez}
Thus the internal degrees of freedom of both the impurity and the conduction electron system
are important for the Kondo effect in real materials.

\acknowledgements
We are grateful to S. Curnoe, H. Kusunose, O. Sakai, H. Shiba, R. Shiina, and K. Ueda
for helpful discussions and comments.

\appendix
\section{Another Derivation of the NRG Hamiltonian}
In Appendix, we present another derivation of the NRG Hamiltonian (\ref{eqn:NRG}),
following Satori {\it et al.}.
\cite{Satori}
First, we apply the logarithmic discretization to the conduction band.
Then we obtain the hopping type of Hamiltonian,
\begin{eqnarray}
{\bar H}_{\rm kin} &=& {1 + \Lambda^{-1} \over 2} \cr
&\times& \sum_{n =0}^{\infty} \sum_{l\sigma}
\Lambda^{-n/2} \varepsilon_n (f_{nl\sigma}^{\dag} f_{n + 1,l\sigma}
 + f_{n + 1,l\sigma}^{\dag} f_{nl\sigma}),
\end{eqnarray}
for the free electrons.
Here $f_{nl \sigma}^{\dag}$ and $f_{nl \sigma}$ are fermion operators
corresponding
to the conduction electrons with angular momentum $l$.
For the BCS pairing interaction in the $p_x + i p_y$-wave superconductors,
we can also obtain the following Hamiltonian:
\begin{eqnarray}
{\bar H}_{\Delta} = \sum_{n = 0}^{\infty} \sum_{\sigma}
(-i {\bar \Delta}) (f_{n,0,\sigma}^{\dag} f_{n,1,-\sigma}^{\dag}
 - f_{n,1,-\sigma} f_{n,0,\sigma}).
\label{eqn:H-delta-p}
\end{eqnarray}
To diagonalize ${\bar H}_{\Delta}$, we use the following Bogoliubov
transformation:
\begin{mathletters}
\begin{eqnarray}
&& b_{n,+,\sigma}^{\dag} = {1 \over \sqrt{2}} (f_{n,0,\sigma}^{\dag} - i
f_{n,1,-\sigma}), \\
&& b_{n,-,-\sigma} = {1 \over \sqrt{2}} (f_{n,1,-\sigma} - i
f_{n,0,\sigma}^{\dag}),
\end{eqnarray}
\end{mathletters}\noindent
and we obtain
\begin{eqnarray}
{\bar H}_{\Delta} = -{\bar \Delta} \sum_{n = 0}^{\infty}
\left(\sum_{\tau \sigma} b_{n \tau \sigma}^{\dag} b_{n \tau \sigma} - 2 \right).
\end{eqnarray}
The impurity Hamiltonian is expressed by the Bogoliubov operators
$b_{n \tau \sigma}^{\dag}$ and $b_{n \tau \sigma}$ as
\begin{eqnarray}
&& {\bar H}_{\rm imp} = - \bar{J} \sum_{\sigma \sigma'} \bS \cdot \bsigma_{\sigma \sigma'}
f_{00 \sigma}^{\dag} f_{00 \sigma'} \nonumber \\
&& ~~~~= -{\bar{J} \over 2} \sum_{\sigma \sigma'} \bS \cdot \bsigma_{\sigma \sigma'}
(b_{0,+,\sigma}^{\dag} + i b_{0,-,-\sigma}) \cr
&&~~~~~~~~~~~~~~~~~~~~~~~~\times(b_{0,+,\sigma'} - i b_{0,-,-\sigma'}^{\dag}).
\end{eqnarray}
The Hamiltonian for the conduction electrons is given by
\begin{eqnarray}
{\bar H}_{\rm kin} &=& {1 + \Lambda^{-1} \over 2} \sum_{n =0}^{\infty} \sum_{\sigma}
\Lambda^{-n/2} \varepsilon_n (-i) \cr
&\times& (b_{n,+,\sigma} b_{n+1,-,-\sigma}+ b_{n,+,\sigma}^{\dag} b_{n+1,-,-\sigma}^{\dag} \cr
&& + b_{n+1,+,\sigma} b_{n,-,-\sigma}
   + b_{n+1,+,\sigma}^{\dag} b_{n,-,-\sigma}^{\dag}).
\end{eqnarray}
Next we apply the following particle-hole transformation:
\begin{eqnarray}
&& c_{2n,+,\sigma}^{\dag} = b_{2n,+,\sigma}^{\dag},~~
c_{2n,-,\sigma}^{\dag} = i b_{2n,-,-\sigma}, \nonumber \\
&& c_{2n-1,+,\sigma}^{\dag} = -i b_{2n-1,-,-\sigma},~~
c_{2n-1,-,\sigma}^{\dag} = -b_{2n-1,+,\sigma}^{\dag}.
\end{eqnarray}
In this manner we can exchange the roles of particles and holes to keep the number
of fermion quasiparticles.
Then we obtain the same form of the Hamiltonian given in (\ref{eqn:NRG}).
The above argument can be applied to the case of $d_{x^2 - y^2} + i
d_{xy}$-wave
superconductors by replacing $i \Delta$ to $-\sigma \Delta$.
In this case, the Bogoliubov transformation is given by
\begin{mathletters}
\begin{eqnarray}
&& b_{n,+,\sigma}^{\dag} = {1 \over \sqrt{2}}(f_{n,0,\sigma}^{\dag} +
\sigma f_{n,2,-\sigma}), \\
&& b_{n,-,-\sigma} = {1 \over \sqrt{2}} (f_{n,2,-\sigma} - \sigma
f_{n,0,\sigma}^{\dag}),
\end{eqnarray}
\end{mathletters}\noindent
and the roles of particles and holes are exchanged by the following
transformation:
\begin{eqnarray}
&& c_{2n,+,\sigma}^{\dag} = b_{2n,+,\sigma}^{\dag},~~
c_{2n,-,\sigma}^{\dag} = -\sigma b_{2n,-,-\sigma}, \nonumber \\
&& c_{2n-1,+,\sigma}^{\dag} = -\sigma b_{2n-1,-,-\sigma},~~
c_{2n-1,-,\sigma}^{\dag} = b_{2n-1,+,\sigma}^{\dag}.
\end{eqnarray}

\vspace{-8mm}



\end{document}